\documentclass[referee,pdflatex,sn-mathphys-num]{sn-jnl}

\usepackage{graphicx}%
\usepackage{multirow}%
\usepackage{amsmath,amssymb,amsfonts}%
\usepackage{amsthm}%
\usepackage{mathrsfs}%
\usepackage[title]{appendix}%
\usepackage{xcolor}%
\usepackage{textcomp}%
\usepackage{manyfoot}%
\usepackage{booktabs}%
\usepackage{algorithm}%
\usepackage{algorithmicx}%
\usepackage{algpseudocode}%
\usepackage{listings}%
\usepackage{siunitx}%
\usepackage{hyperref}%

\theoremstyle{thmstyleone}%
\theoremstyle{thmstyletwo}%

\theoremstyle{thmstylethree}%

\begin{document}

\title[Article Title]{Near-field Dressing of Thermal Emission}


\author[1]{\fnm{Victor} \sur{Guillemot}}

\author[1]{\fnm{Raphael} \sur{Mol}}

\author[3]{\fnm{Riccardo} \sur{Messina}}

\author[1]{\fnm{Valentina} \sur{Krachmalnicoff}}

\author[3]{\fnm{Philippe} \sur{Ben-Abdallah}}

\author*[2]{\fnm{Wilfrid} \sur{Poirier}}\email{wilfrid.poirier@lne.fr}

\author*[1]{\fnm{Yannick} \sur{De Wilde}}\email{yannick.dewilde@espci.fr}

\affil*[1]{\orgdiv{Institut Langevin, ESPCI Paris, Université PSL, CNRS, 75005 Paris, France}}

\affil[3]{\orgdiv{Laboratoire Charles Fabry, UMR 8501, Institut d'Optique, CNRS, Universit\'{e} Paris-Saclay, 91127 Palaiseau Cedex, France}}

\affil[2]{\orgdiv{Laboratoire national de métrologie et d'essais (LNE), 29 avenue Roger Hennequin, 78197 Trappes, France}}


\abstract{
Radiative heat transfer at subwavelength distances is generally understood as enhanced energy exchange mediated by photon tunnelling between neighboring bodies. While near-field interactions can dramatically increase mutual heat transfer, whether they also modify the thermal radiation emitted by the bodies themselves remains an open question. Here we experimentally show that near-field electromagnetic coupling reshapes far-field thermal emission through a distance-dependent dressed emissivity. Using a dual-probe calorimetric platform, we independently monitor the radiative balance of two borosilicate microspheres over separations ranging from \SI{120}{\micro\meter} to a few hundred nanometers, spanning the transition from the far field to the near field. Nanowatt-resolved differential radiometry reveals asymmetric heat fluxes and a non-monotonic response of the hotter sphere, demonstrating that thermal radiation is governed not only by emitter–bath interactions but also by coupling to the surrounding photonic environment. By analysing the total power exchanged between the coupled system and the external thermal bath, we directly extract a dressed emissivity and show that near-field interactions renormalize the far-field thermal emission of the pair through a redistribution of the electromagnetic modes available to thermal fluctuations. These observations provide direct experimental evidence that thermal emitters are dressed by their electromagnetic environment, establishing a thermal analogue of the Purcell effect.
}


\maketitle


Radiative heat transfer originates from the electromagnetic field generated by thermally fluctuating currents~\cite{polder_theory_1971}. When two objects are separated by distances comparable to or smaller than the thermal wavelength, 
evanescent electromagnetic modes can tunnel across the gap and produce heat fluxes that exceed the 
blackbody limit~\cite{joulain_definition_2003, mulet_enhanced_2002}. This near-field regime has been extensively investigated in planar and tip–surface geometries, establishing the fundamental mechanisms governing nanoscale thermal radiation and highlighting its potential for thermal management and energy conversion~\cite{rousseau_radiative_2009, narayanaswamy_near-field_2008, fiorino_giant_2018, yan_surface_2023,kittel_near-field_2005, mittapally_probing_2023,latella_graphene-based_2021,song_thermophotovoltaic_2022, mittapally_near-field_2021, bhatt_integrated_2020,lucchesi_near-field_2021}. 
More generally, electromagnetic emission is known to depend strongly on the surrounding photonic environment. In quantum electrodynamics, Purcell predicted that the spontaneous emission rate of an emitter is controlled by the electromagnetic density of states of its environment. This concept was later demonstrated experimentally in the pioneering works of Drexhage and Chance~\cite{drexhage_influence_1970,chance}. These studies established that radiative properties, including the fluorescence lifetime and the emitted power, are not intrinsic to emitters alone but emerge from their coupling to the surrounding electromagnetic field~\cite{carminati_electromagnetic_2015}.

In fluctuational electrodynamics, thermal emission similarly originates from fluctuating dipolar sources and should therefore also depend on the electromagnetic environment. In analogy with the Purcell effect, nearby objects can modify the electromagnetic modes available to thermal fluctuations and thereby renormalize the radiative properties of thermal emitters. Theoretical works have predicted environment-mediated phenomena such as long-range heat transfer~\cite{asheichyk_long-range_2023}, collective thermal transport~\cite{muller_many-body_2017} and enhancement or suppression of thermal radiation through many-body electromagnetic coupling~\cite{ song_many-body_2021,messina_three-body_2012, ben-abdallah_multitip_2019,abou-hamdan_transition_2022,abou-hamdan_hybrid_2021}, leading to non-monotonic radiative behaviors~\cite{guillemot_nonmonotonic_2025}.
Despite these predictions, direct experimental evidence that the electromagnetic environment modifies the thermal emission of coupled emitters remains scarce, especially in fully three-dimensional geometries where objects simultaneously exchange radiation with each other and with the surrounding thermal bath. While experiments have demonstrated substantial enhancements of heat transfer between neighboring subwavelength bodies, together with profound modifications of their spectral coupling characteristics~\cite{thompson_hundred-fold_2018, tang_corner_2024}, they have not directly investigated how the environment governs the external radiative channels of the interacting system.

Here we experimentally investigate radiative heat transfer between two borosilicate microspheres while independently monitoring the temperature of each sphere. The spheres are mounted on a unique dual-probe platform combining sub-milliKelvin thermometry with nanowatt-scale calorimetric resolution, and are separated by gaps ranging from \SI{120}{\micro\meter} down to a few hundred nanometers, spanning the transition from the far-field to the near-field regime. 
Our measurements reveal pronounced asymmetries in the radiative response of the two spheres together with a non-monotonic evolution of the heat flux at short separations. By analyzing the total power exchanged between the coupled system and the surrounding bath, we isolate the contribution associated with the external radiative channels and introduce a distance-dependent dressed emissivity that quantifies how the photonic environment modifies thermal emission. Comparison with full-wave fluctuational electrodynamics calculations shows that these effects originate from near-field electromagnetic coupling, which reshapes the spectrum and efficiency of thermal radiation emitted by the coupled system.

\clearpage
\section{Results}

\begin{figure}[!h]
    \centering
    \includegraphics[width = 1\textwidth]{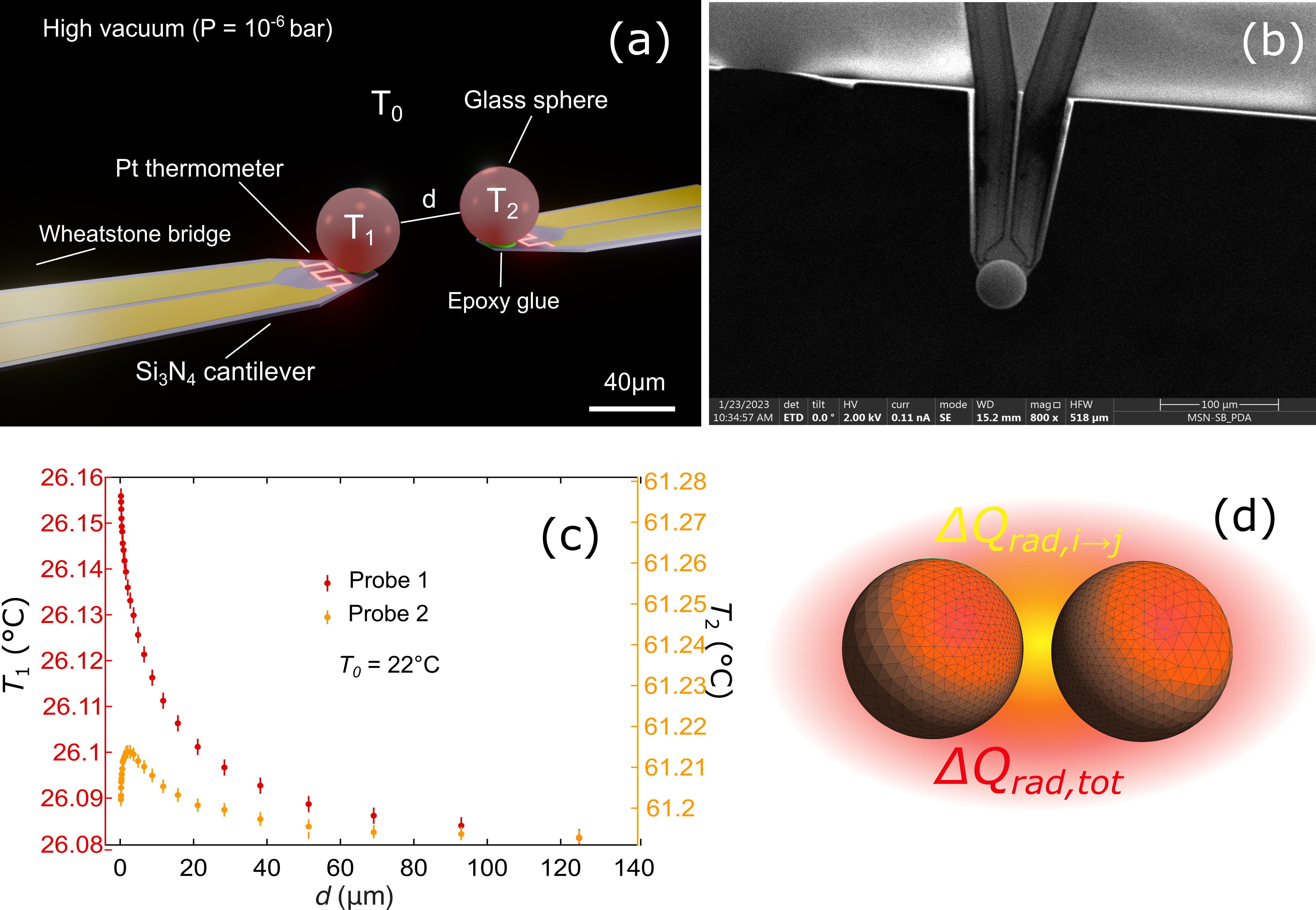}
    \caption{\textbf{Dual-probe platform for resolving individual radiative balances in a coupled thermal dimer.} 
    (a) Schematic of the experimental configuration. Two \SI{42.3}{\micro\meter}-diameter borosilicate microspheres are attached to the apices of opposing scanning thermal microscopy probes and separated by a controllable gap $d$. Each cantilever integrates a platinum resistive thermometer providing both local temperature readout ($T_1$, $T_2$) and Joule heating through an injected electrical power $P_\mathrm{Joule}$. The experiment is performed under high vacuum ($P=\SI{e-6}{\milli\bar}$) to suppress heat transfer by conduction and convection through air. Each sphere exchanges thermal radiation both with the neighboring sphere and with the surrounding environment at temperature $T_0$. This unique dual-probe configuration provides simultaneous access to the local radiative balance of each emitter, enabling direct investigation of how electromagnetic coupling modifies thermal emission.
    (b) Top-view scanning-electron micrograph of one probe showing the borosilicate microsphere and the gold tracks used to electrically connect the integrated platinum thermometer.
    (c) Simultaneous temperature variations of the two spheres as a function of separation. As the gap decreases, the colder sphere exhibits a monotonic temperature increase, whereas the hotter sphere displays a non-monotonic response. This asymmetry reveals the presence of three interacting thermal reservoirs ($T_0$, $T_1$ and $T_2$), so that each sphere exchanges radiation both with its counterpart and with the surrounding electromagnetic bath.
    (d) Meshed geometry used for fluctuational electrodynamics calculations with the boundary-element solver SCUFF-EM. The mesh is refined in the region where the two spheres face each other to accurately resolve near-field coupling. The modulated sphere--sphere radiative exchange is denoted $\Delta Q_{\mathrm{rad},i\rightarrow j}$, whereas $\Delta Q_{\mathrm{rad,tot}} = \sum_{i=1}^2 \Delta Q_{\mathrm{rad},i}$ denotes the net radiative exchange between the coupled system and the surrounding bath. Separating these two contributions enables direct access to the external radiative channels of the interacting system and forms the basis for extracting the dressed emissivity introduced in this work.
    }
    \label{fig:1}
\end{figure}


Our experimental platform consists of two identical scanning thermal microscopy probes facing each other, each terminated by a borosilicate microsphere and equipped with an integrated platinum thermistor (Fig.~\ref{fig:1}a,b). The probes operate under high vacuum ($P=\SI{e-6}{\milli\bar}$) to suppress heat transfer by conduction and convection through air, while the sphere--sphere separation $d$ is controlled with nanometer resolution (see Online Methods~\ref{plateform}). The thermistors provide both local temperature readout through resistance measurement using metrological Wheatstone bridge~\cite{Poirier2026ResistanceBridge} and lockin-detection, enabling independent control and simultaneous monitoring of both sphere temperature (see online Methods ~\ref{plateform}). The Joule power injected on each probe is given by $P_\mathrm{Joule} = I^2 R$, where $I$ is the injected current and $R$ the resistance of the thermistor, which depends linearly on the temperature. The dual-probe architecture provides simultaneous access to the local radiative balance of each emitter, enabling nanowatt-resolved measurements of heat exchange while independently controlling the temperature of both spheres.

Unlike conventional near-field heat-transfer experiments, which typically measure only the net power exchanged between two bodies, our platform independently probes the radiative response of each emitter. This capability is essential for disentangling sphere–sphere exchange from radiation exchanged with the surrounding electromagnetic bath. Figure~\ref{fig:1}(c) shows an approach curve obtained while reducing the separation between the two spheres. Because the thermistors on the two probes are read independently, the temperatures of both spheres can be monitored simultaneously as the gap changes. The two temperatures do not evolve symmetrically despite the geometric symmetry of the system. In an environment at $T_0=\SI{22}{\celsius}$, the colder sphere ($T_1\approx\SI{26}{\celsius}$) exhibits a monotonic temperature increase as it approaches the hotter one ($T_2\approx\SI{61}{\celsius}$). By contrast, the hotter sphere shows a non-monotonic response: its temperature first increases at large distances but then decreases sharply once the separation becomes comparable to the thermal wavelength ($\lambda_\mathrm{Wien}\approx\SI{10}{\micro\meter}$). The statistical uncertainty of the temperature measurement, determined from repeated measurements, is approximately \SI{1}{\milli\kelvin}  and is represented by the error bars. This asymmetric temperature evolution indicates that the two spheres experience different radiative balances, reflecting the presence of three thermal reservoirs ($T_0$, $T_1$ and $T_2$).

To quantify these observations, we convert the measured resistance variations $\Delta R_i$ into a spatially modulated radiative heat flux such that $\Delta Q_{\mathrm{rad},i}(d)=k_i\,\Delta R_i(d)\,G_i $ for each sphere ($i=1,2$), where $\Delta $ refers to a difference between the flux measured at distance $d$ and the one measured at a reference distance in far-field $d_\mathrm{ref}=$\SI{120}{\micro\meter}, $k_i$ is the sensitivity expressed in \SI{}{\kelvin\per\ohm} and $G_i$ is the thermal conductance of the probe (see Online Methods~\ref{modulation}). These experimental flux variations are compared with fluctuational-electrodynamics calculations performed with SCUFF-EM, a boundary-element Maxwell solver that computes radiative heat transfer in arbitrary three-dimensional geometries~\cite{SCUFF1,SCUFF2}. The numerical geometry used in the calculations is shown in Fig.~\ref{fig:1}(d), where the mesh is refined in the region where the two spheres face each other in order to accurately resolve near-field coupling. In this representation, $\Delta Q_{\mathrm{rad}, i \rightarrow j}$ denotes the modulated radiative exchange between the two spheres, while $\Delta Q_{\mathrm{rad,tot}} = \sum_{i=1}^2 \Delta Q_{\mathrm{rad},i}$ represents the total radiative exchange between the coupled two-sphere system and the surrounding environment.

\begin{figure}[!h]
    \centering
    \includegraphics[width = 1\textwidth]{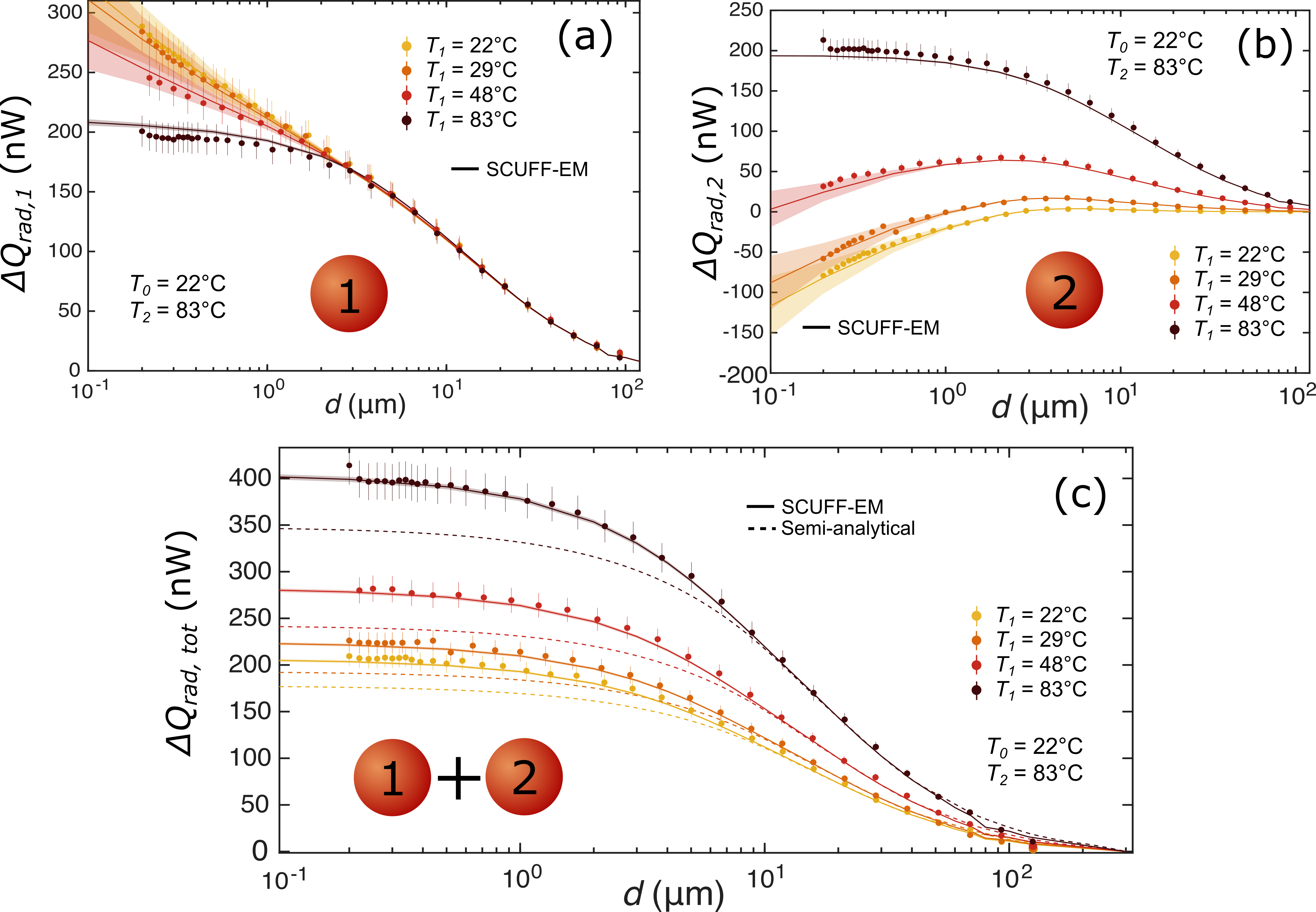}
    \caption{\textbf{Direct observation of asymmetric radiative balances in near-field-coupled thermal emitters}
    (a) Variation of radiative heat flux of the sphere~1, $\Delta Q_{\mathrm{rad},1}(d)$, measured for four different temperatures $T_1$ while sphere~2 is maintained at $T_2=\SI{83}{\celsius}$ and the environment at $T_0=\SI{22}{\celsius}$. Experimental data are compared with full-wave fluctuational-electrodynamics calculations performed with SCUFF-EM (solid lines). In the far field, the radiative flux of sphere~1 depends weakly on $T_1$, whereas at short separations it becomes strongly temperature dependent due to the onset of near-field coupling between the two spheres. 
    (b) Variation of the radiative heat flux of the sphere~2, $\Delta Q_{\mathrm{rad},2}(d)$. In contrast with sphere~1, the hotter sphere exhibits a non-monotonic dependence on distance: the flux initially increases as the spheres approach, reflecting a redistribution of propagating exchange with the bath, but decreases at short separations once evanescent sphere–sphere coupling dominates. 
    (c) Total radiative exchange between the coupled two-sphere system and the surrounding environment, $\Delta Q_{\mathrm{rad,tot}}=\Delta Q_{\mathrm{rad},1}+\Delta Q_{\mathrm{rad},2}$. Because the mutual sphere–sphere exchange cancels by energy conservation, this quantity provides direct experimental access to the external radiative channels of the coupled system. Solid lines show SCUFF-EM calculations and dashed lines the semi-analytical model described in the Online Methods.}
    \label{fig:2}
\end{figure}

To explore the different thermal configurations, four independent experiments were performed while varying the temperature of sphere~1 by adjusting the Joule heating power, while sphere~2 was maintained at $T_2\approx\SI{83}{\degreeCelsius}$ and the environment at $T_0=\SI{22}{\degreeCelsius}$. The resulting radiative fluxes are shown in Fig.~\ref{fig:2}. Figures~\ref{fig:2}(a,b) display the modulated radiative heat flux  $\Delta Q_{\mathrm{rad},1}(d)$ and $\Delta Q_{\mathrm{rad},2}(d)$, received by sphere~1 and sphere~2, respectively, as a function of the separation $d$. The symbols correspond to experimental measurements, while the solid lines show the SCUFF-EM predictions. The shaded regions represent the numerical uncertainty associated with the experimental uncertainty on the gap distance, which becomes more significant at short separations.

The measurements show excellent agreement with the SCUFF-EM calculations over the entire explored range, from  \SI{120}{\micro\meter} down to a few hundred nanometers, for all four temperature configurations. In the nearly symmetric case $T_1 \approx T_2 \approx \SI{83}{\degreeCelsius}$ (dark curves in Fig.~\ref{fig:2}(a,b)), the two spheres behave identically, in fact, reciprocity imposes $\Delta Q_{\mathrm{rad},1}=\Delta Q_{\mathrm{rad},2}$. In this configuration, the radiative flux tends to a distance-independent value at short separations since no net near-field exchange occurs between the two spheres.

Away from this symmetric situation, the radiative responses of the two spheres become markedly different. For the colder sphere, the far-field behavior is primarily governed by a geometric redistribution of propagating radiative exchange between the bath and the neighboring sphere. As the separation decreases further, below the thermal wavelength, an additional contribution appears, leading to an increase of the flux through evanescent near-field coupling with the hotter sphere. By contrast, the hotter sphere exhibits a non-monotonic dependence on distance: its radiative flux initially increases as the gap decreases, but then decreases sharply once near-field exchange between the spheres dominates.

To interpret these trends, we introduce a compact semi-analytical description in which each sphere exchanges radiation both with the thermal bath and with the other sphere. Exact calculations of radiative heat transfer between two spheres have previously been obtained within fluctuational electrodynamics \cite{sasihithlu_proximity_2011,narayanaswamy_thermal_2008}. However, these approaches describe only the mutual exchange between the two objects and do not explicitly separate the contribution associated with radiation exchanged with the surrounding environment. In our experiment, each sphere interacts simultaneously with the other sphere and with the thermal bath, so that both channels must be considered. The net radiative power received by sphere~$i$ can then be written as
\begin{equation}
    Q_{\mathrm{rad},i}(d)=Q_{\mathrm{bath} \rightarrow i}^{\mathrm{FF}}(d)+Q_{j \rightarrow i}^{\mathrm{FF}}(d)+Q_{j \rightarrow i}^{\mathrm{NF}}(d),
\end{equation}
where the three terms represent far-field exchange with the bath, far-field exchange with the other sphere, and near-field exchange between the two spheres.

In Landauer form, the radiative flux reads
\begin{multline}
    Q_{\mathrm{rad},i}(d)=\int_{0}^{\infty}\frac{\mathrm{d}\omega}{2\pi}\Big[
    \Delta\Theta(\omega,T_i,T_0)\,\mathcal{T}_{\mathrm{bath} \rightarrow i}^{\mathrm{FF}}(\omega,d) \\
    +\Delta\Theta(\omega,T_i,T_j)\,\mathcal{T}_{j\rightarrow i}^{\mathrm{FF}}(\omega,d)
    +\Delta\Theta(\omega,T_i,T_j)\,\mathcal{T}_{j\rightarrow i}^{\mathrm{NF}}(\omega,d)\Big],
\end{multline}
where $\Delta\Theta(\omega,T_i,T_j)=\Theta(\omega,T_j)-\Theta(\omega,T_i)$ and $\Theta(\omega,T)=\hbar\omega/(e^{\hbar\omega/k_\mathrm{B}T}-1)$. The explicit expressions of the transmission coefficients are given in the Online Methods.

This formulation highlights the origin of the observed asymmetry. The exchange with the bath is weighted by $\Delta\Theta(\omega,T_i,T_0)$, which is always a non-negative quantity in our experiments, whereas the sphere--sphere exchange depends on $\Delta\Theta(\omega,T_i,T_j)$, which can be either positive, negative or zero. As soon as $T_1\neq T_2$, these two contributions cannot be related by symmetry, so the local radiative balances of the two spheres differ, and a competition emerges between radiation exchanged with the surrounding bath and radiation exchanged with the neighboring sphere. Indeed, the surrounding environment acts as an active third body in the energy exchange.

The same decomposition also explains the non-monotonic response observed for the hotter sphere. At large separations, reducing the gap primarily redistributes propagating exchange with the bath. At shorter distances, however, evanescent coupling between the spheres becomes dominant and drives the radiative flux in the opposite direction. The semi-analytical model captures these qualitative trends (see Online Methods~\ref{semi-analytical}), while the full SCUFF-EM calculations are required to reproduce quantitatively the behavior at the smallest separations where multiple scattering and wave effects become important.

This three-body description naturally leads to considering the total radiative exchange between the coupled two-sphere system and the surrounding environment. While the mutual exchange between the spheres dominates the near-field heat transfer, the external emission of the pair toward the bath provides direct information on how the electromagnetic environment modifies the radiative properties of each emitter. In particular, it allows us to quantify how near-field coupling renormalizes the far-field thermal emission of the dimer of spheres.

Figure~\ref{fig:2}(c) shows the total modulated radiative exchange with the environment, $\Delta Q_{\mathrm{rad,tot}}=\Delta Q_{\mathrm{rad},1}+\Delta Q_{\mathrm{rad},2}$. This quantity corresponds to the variation of the net radiative emission of the two-sphere system toward the thermal bath at temperature $T_0$. Because the mutual radiative exchange between the spheres conserves energy, the sphere–sphere contributions cancel in this sum, leaving only the external radiation exchanged with the bath.
As the separation decreases, $\Delta Q_{\mathrm{rad,tot}}$ increases, indicating that the coupled system emits progressively less energy toward the environment. Indeed, as the spheres approach each other, a larger fraction of the electromagnetic energy is exchanged between the two objects rather than radiated into the far field.
The dashed line shows the prediction of the semi-analytical model based solely on the view-factor description (see Online Methods~\ref{semi-analytical}). This model accurately reproduces the behavior at large separations, where radiative exchange is dominated by geometrical far-field considerations. However, it progressively deviates from the experimental data as the gap becomes comparable to the thermal wavelength, revealing the onset of wave effects associated with near-field electromagnetic coupling.

\begin{figure}[!h]
    \centering
    \includegraphics[width = 0.96\textwidth]{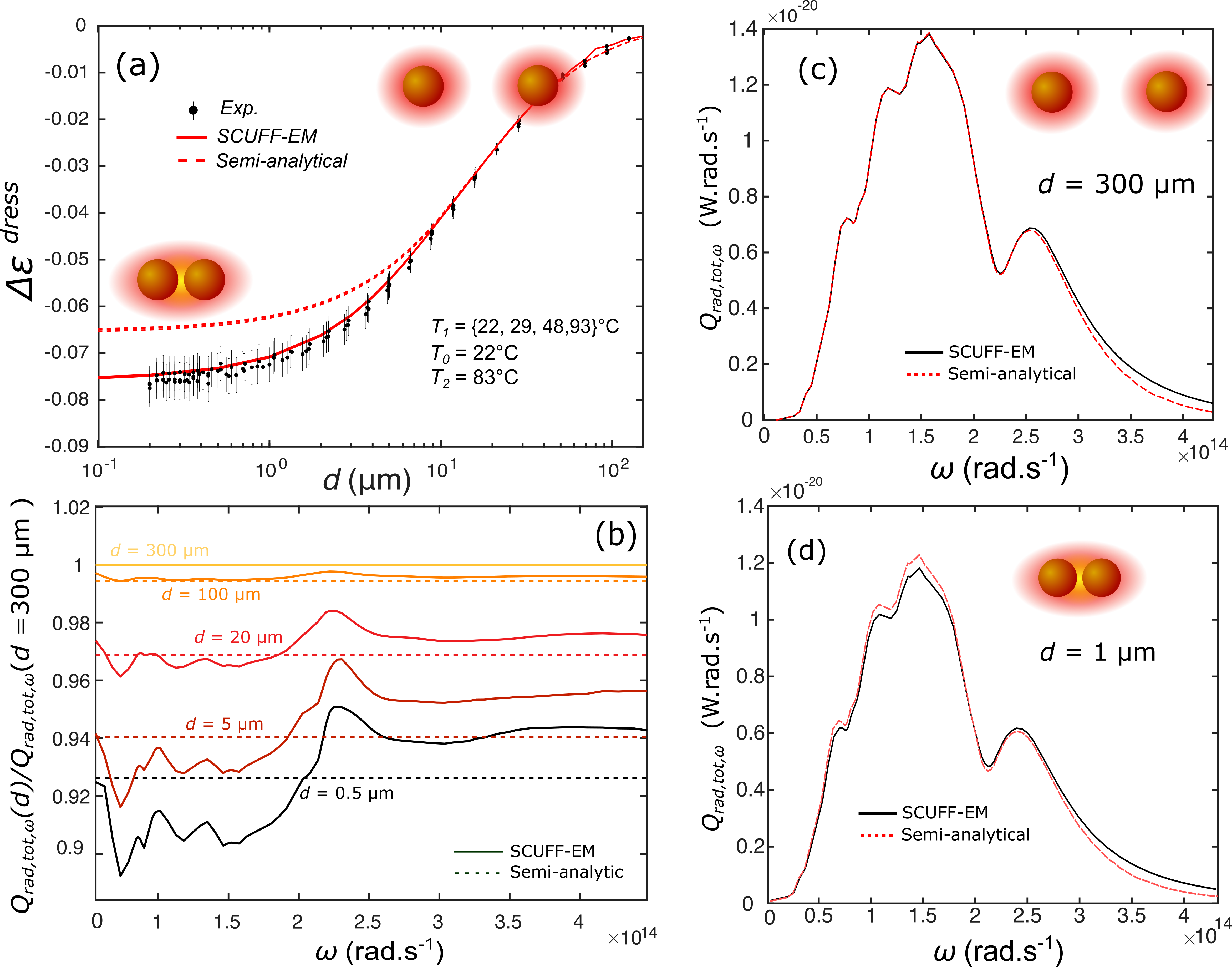}
    \caption{\textbf{Renormalization of far-field emission by near-field coupling.}
    (a) Distance-dependent dressed emissivity of the coupled two-sphere system, defined from the total radiative exchange with the environment. Experimental measurements (for all values of $T_1$) are represented with dots, calculations performed with SCUFF-EM are represented as solid lines and the semi-analytical model described in the Online Methods is represented as dashed lines. As the spheres approach each other in the far-field, the dressed emissivity decreases, as predicted by the semi-analytical model. At short distance, deviations from the geometrical model indicate that near-field coupling modifies the far-field emission of the pair. The collapse of all datasets onto a single curve demonstrates that the emissivity modification is governed primarily by the electromagnetic coupling between the spheres and is essentially independent of the specific temperature configuration. (b) Spectral radiative emission of the two-sphere system calculated with the semi-analytical model (dashed lines) and with SCUFF-EM (solid lines), normalized by the spectrum at $d=\SI{300}{\micro\meter}$. The SCUFF-EM calculations reveal a distance-dependent redistribution of the emitted spectrum as the spheres approach, reflecting the modification of the electromagnetic modes participating in thermal radiation. In contrast, in the semi-analytical model, the normalized spectrum is simply rescaled by a constant factor and therefore remains wavelength-independent, distance-dependent modifications of the electromagnetic modes cancels out in the model. (c,d) Absolute radiative spectra predicted by the semi-analytical model (dashed line) and by SCUFF-EM (solid line) for representative separations. At large distances (c) the spectra approach the emission of two isolated spheres, whereas at short separations (d)coupling-induced spectral reshaping appears. These results demonstrate that near-field interactions between the spheres not only affect their mutual heat exchange but also renormalize the far-field thermal emission of the dimer of spheres.}
    \label{fig:3}
\end{figure}

To quantify how the electromagnetic environment modifies the thermal emission of the two-sphere system, we introduce the dressed emissivity variation, $\Delta \varepsilon^{\mathrm{dress}}$. This quantity measures the change in far-field thermal radiation power of the coupled pair relative to that of two isolated spheres (see Online Methods~\ref{dressed}). It is defined from the total radiative exchange with the environment as
\begin{equation}
\Delta \varepsilon^{\mathrm{dress}}(d)=\varepsilon^{\mathrm{dress}}(d)-\varepsilon^{\mathrm{sphere}}
= \frac{\Delta Q_{\mathrm{rad,tot}}(T_0,T_1,T_2;d)}
{A\sigma\left(2T_0^4-T_1^4-T_2^4\right)},
\end{equation}
where $A=4\pi R^2$ is the surface area of a sphere and $\sigma$ is the Stefan--Boltzmann constant. This definition isolates the modification of thermal emission induced by the presence of a nearby object through near-field electromagnetic interactions.

The dressed emissivity variation is closely related to the thermal Purcell factor. Indeed, the Purcell factor is commonly defined as the ratio between the power emitted in the presence of an electromagnetic environment and that emitted in free space (see Supplementary Information). In our experimental configuration, however, differential radiometric measurements provide direct access to the quantity $\Delta \varepsilon^{\mathrm{dress}}(d)$ rather than the Purcell factor itself. 

Figure~\ref{fig:3}(a) shows the distance dependence of this dressed emissivity. Remarkably, all datasets collapse onto a single curve, indicating that the dressed emissivity is primarily governed by the electromagnetic coupling between the spheres rather than by the specific temperature configuration. This weak temperature dependence indicates that the effect is predominantly electromagnetic in origin. It originates from the modification of the photonic environment surrounding each emitter.

At large distances, the quantity $\Delta \varepsilon^{\mathrm{dress}}$ tends to zero, indicating that the system composed of the two-sphere behaves as two isolated spheres with an emissivity given by Mie theory. As the spheres approach each other, the variation of dressed emissivity decreases and reaches a minimum at the smallest gaps, reflecting the net reduction of the far-field emission of the pair. Full-wave calculations performed with SCUFF-EM quantitatively reproduce this reduction. On the other hand, the semi-analytical model captures the qualitative trend but systematically underestimates the magnitude of the effect. This difference highlights the physical origin of the emissivity modification. At large separations, the radiative exchange is well described by a simple view-factor picture in which each sphere emits almost independently toward the environment. In this regime the photonic environment remains essentially unchanged and both models coincide. To understand the difference between the two models, one can analyze the spectral dependencies of the emitted radiation. 

Figure~\ref{fig:3}(b) shows the radiative emission spectrum of the coupled two-sphere system, normalized by the spectrum obtained at large separation. In the semi-analytical model, the spectral modification arises purely from geometric considerations associated with the view-factor description, since the internal heat fluxes $\Delta Q_{\mathrm{i\rightarrow j}}$ exchanged between the two spheres cancel out in this description. As a consequence, the normalized spectrum is just rescaled by a constant factor and remains independent of wavelength.

By contrast, the full-wave SCUFF-EM calculations reveal clear wavelength-dependent spectral modifications as the spheres approach each other. In particular, a distinct spectral feature appears around $\omega \approx 2.2\times10^{14}\,\mathrm{rad\,s^{-1}}$, corresponding to the Mie resonance of a single sphere, showing the modification of this resonance for the coupled system. This spectral behavior indicates that geometry alone cannot account for the observed emissivity renormalization. Instead, the wavelength-dependent reshaping reflects wave effects associated with near-field electromagnetic coupling, which modifies the photonic environment and the electromagnetic modes available for thermal radiation.

Figures~\ref{fig:3}(c) and~\ref{fig:3}(d) compare the emission spectra of the two-sphere system predicted by the semi-analytical model and by SCUFF-EM at two different distances. At large separations (Figures~\ref{fig:3}(c)), the agreement between the two calculations is excellent, indicating that the thermal emission of the system can be accurately described by considering two independent spheres with Mie-theory emissivities. At short separations, however, Fig.~\ref{fig:3}(d) systematic deviations emerge, revealing the onset of coupling-induced modifications of the electromagnetic modes participating in thermal radiation. These deviations originate from the modification of the local electromagnetic density of states around each sphere and account for the observed emissivity renormalization. We use the term thermal analogue of the Purcell effect because the variation of emitted power results from a change in the electromagnetic modes available to thermal fluctuations, in direct analogy with the environment-controlled spontaneous emission of quantum emitters.

\section{Discussion}

Our measurements establish the sphere--sphere--bath configuration as a minimal experimental platform to probe many-body radiative heat transfer in a fully three-dimensional geometry. The central result is not merely the enhancement of heat transfer at short separations, but the emergence of asymmetric radiative balances for the two spheres as soon as their temperatures differ. This asymmetry cannot be captured within a closed two-body description and directly reveals the role of the surrounding electromagnetic bath as an active participant in the energy exchange.

Beyond the present configuration, our results highlight the broader role of the photonic environment in shaping thermal emission. In analogy with emitter dressing in nano-optics, the presence of nearby objects modifies the electromagnetic modes available to thermal fluctuations and thereby renormalizes the emissivity of the coupled system. Similarly to the Purcell factor, the quantity $\Delta\varepsilon^{\mathrm{dress}}$ provides a direct experimental measure of how near-field interactions alter the far-field radiation properties of thermal emitters.

More generally, our work bridges the conventional separation between near-field heat transfer and far-field thermal emission. While these two phenomena are often treated independently, our results show that they are intrinsically linked through the electromagnetic environment. The present experiments reveal an extra reduction of the dressed emissivity as the spheres approach each other that cannot be explained by geometrical considerations, indicating that near-field interactions can suppress the radiative outcoupling of thermal fluctuations. However, as discussed in the Supplementary Information, this behavior is not universal. Depending on the nature of the collective electromagnetic modes and their radiative efficiency, near-field coupling may either inhibit or enhance thermal emission. Future investigations of resonant plasmonic and polaritonic systems may therefore provide access to both suppressed and enhanced dressed-emissivity regimes leading to $\Delta\varepsilon_{\mathrm{dress}} < 0$ and $\Delta\varepsilon_{\mathrm{dress}} > 0$ respectively (see Supplementary Information), opening the way toward a broader understanding of thermal emitter dressing and environment-controlled thermal radiation.

\section*{Acknowledgements}

This work was supported by the ”Investissements d’Avenir” program launched by the French Government (Labex WiFi) and by the Agence Nationale de la Recherche (NBODHEAT Project No. ANR-21-CE30-0030).

\bibliography{Biblio_1}

\section{Online Methods}

\subsection{Experimental platform and temperature control}
\label{plateform}
\begin{figure}[!h]
    \centering
    \includegraphics[width=\linewidth]{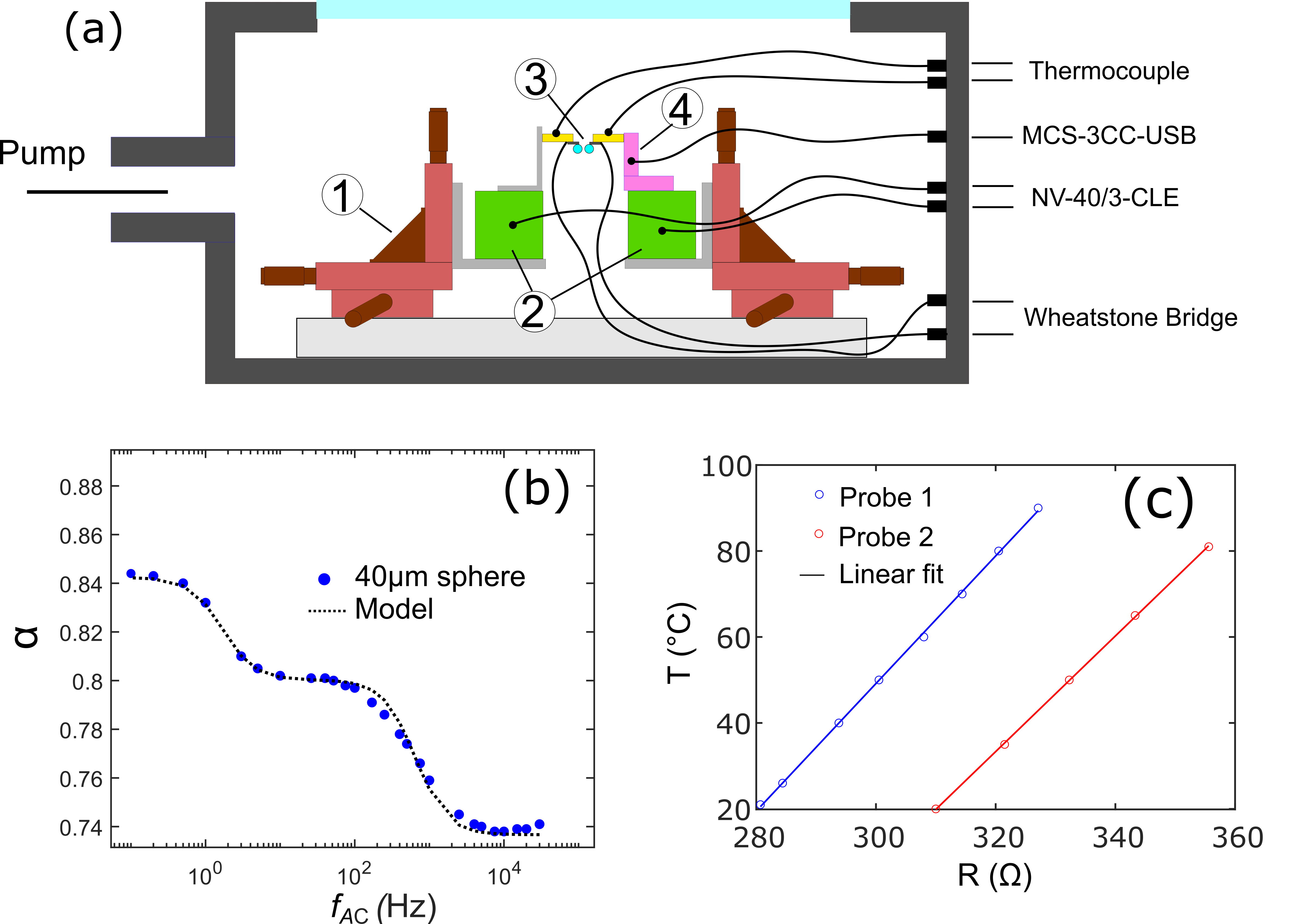}
   \caption{\textbf{Description of the experimental setup.} 
    (a) Schematic of the vacuum chamber. The position of each probe is controlled by a manual micrometric stage combined with a fine piezoelectric stage. 
    (b) Frequency dependence of the bridge response. Two thermal cutoff frequencies are observed. The highest frequency is related to the diffusivity of the cantilever alone, while the lowest comes from the adding of the microsphere. The system is operated above the highest cutoff frequency to avoid temperature oscillations during the measurements.
    (c)Calibration curves of the two probes used in our experiment. The probes are placed in an oven while monitoring their electrical resistance. The temperature coefficient is then determined from a linear fit. }
    \label{fig:Sup_setup}
\end{figure}

The experiment uses two identical custom SThM probes facing each other. Borosilicate microspheres of diameter \SI{42.3}{\micro\meter} are glued at the apices of \SI{200}{\micro\meter}-long silicon-nitride cantilevers (Kelvin Nanotechnologies) equipped with platinum thermistors. The surrounding bath is maintained at $T_0=\SI{22}{\degreeCelsius}$ and the setup is operated under high vacuum ($P=\SI{e-6}{\milli\bar}$) to suppress heat transfer by conduction and convection through air. The sphere--sphere distance is controlled by a piezoelectric nanopositioner with \SI{5}{\nano\meter} resolution. A schematic of the vacuum chamber is represented in figure \ref{fig:Sup_setup}(a). The positioning is made from coarse to fine with manual micrometric stages, a stick-slip actuator and a Piezojena nanopositioning system. 

The electrical resistances of the two thermistors are $R_1=\SI{280}{\ohm}$ and $R_2=\SI{310}{\ohm}$ at $T_0=\SI{22}{\degreeCelsius}$. Their temperature coefficients, calibrated in a temperature-controlled oven, are $k_1=\SI{1.48(0.05)}{\kelvin\per\ohm}$ and $k_2=\SI{1.37(0.05)}{\kelvin\per\ohm}$. The calibration is represented on figure \ref{fig:Sup_setup}(c). Two metrological Wheatstone bridges inject a AC current at a frequency  $f_\mathrm{AC} =$ \SI{13}{\kilo\hertz} into each probe and simultaneously read the bridge unbalance voltage with lock-in detection after amplification. This configuration enables independent Joule heating and milliKelvin-resolved local thermometry on both probes.

The metrological resistance bridge~\cite{Poirier2026ResistanceBridge} is an improved version of the prototype described in~\cite{guillemot_nonmonotonic_2025,doumouro_quantitative_2021}. It is a custom Wheatstone bridge enabling measurement of thermometer resistance ranging from \SI{100}{\ohm} and \SI{1000}{\ohm} both in direct current and in alternating current at frequencies up to a few tenths of \SI{}{\kilo \hertz}. It is made of the thermometer resistance, $R_\mathrm{probe}$, an adjustable reference resistance, $R_\mathrm{e}$, ranging from \SI{100}{\ohm} and \SI{1000}{\ohm} and two fixed \SI{100}{\ohm} resistors. It is equipped with three Kelvin arms to achieve an approximate four-wire definition of both $R_\mathrm{probe}$ and $R_\mathrm{e}$, in order to reduce the impact of variation of resistance of leads and connectors. 

Offset, in-phase and quadrature circuits allows fine balance of the bridge. The residual unbalance voltage is detected by a low-noise voltage amplifier (with a base noise of \SI{0.7}{\nano\volt/\sqrt{\hertz}}) before measurement by a lock-in detector. The bridge is symmetrically biased by $ \pm V$ voltage with a grounded midpoint, while the high- and low-potential measurement leads of the resistance thermometer are individually shielded to ground. This reduces circulation of current leakages caused by lead capacitance and therefore increases the frequency bandwidth of the measurement bridge. 

The probe resistance is determined from the bridge imbalance voltage according to
\begin{equation}
R_\mathrm{probe}
=
R_\mathrm{e}
\frac{[1-a(1-2\alpha)]-\gamma\left[\frac{120+bR_\mathrm{e}}{100+R_\mathrm{e}}\right]}
{[1+a(1-2\alpha)]+\gamma\left[\frac{100+bR_\mathrm{e}}{100+R_\mathrm{e}}\right]},
\label{eqn:resistance_probe}
\end{equation}
where $a=2.5\times10^{-2}$, $b=1.1$, $\alpha$ is the potentiometer setting, and $\gamma=\Delta V/V$ is the normalized bridge imbalance voltage. This expression is valid both at equilibrium ($\gamma=0$) and under finite imbalance voltage, allowing direct conversion of the measured voltage into probe resistance. Measurements are performed at an excitation frequency well above the thermal cutoff of the probes (Fig.~\ref{fig:Sup_setup}(b)), whose highest characteristic frequency is approximately \SI{700}{\hertz}. In these conditions, the probe resistance does not oscillate and the expresssion~\ref{eqn:resistance_probe} is valid as in direct current operation. 
%

\subsection{Differential distance-modulation protocol and flux extraction}
\label{modulation}

Because the radiative signals are much smaller than the conductive heat leak through the cantilevers, the experiment is performed in differential mode. For each gap $d$, the thermometer temperatures are measured and rapidly compared with a reference measurement at $d_\mathrm{ref}=\SI{120}{\micro\meter}$~\cite{guillemot_nonmonotonic_2025}. The differential temperature signal is
\begin{equation}
    \Delta T_i=T_i(d)-T_i(d_\mathrm{ref}),
\end{equation}
which strongly suppresses slow drifts of the bath temperature. A representative approach curve is shown in Fig.~\ref{fig:1}(c). The radiative-flux modulation for sphere~$i$ is obtained from the resistance change of its thermometer through
\begin{equation}
    \Delta Q_{\mathrm{rad},i}(d)=k_i\,\Delta R_i(d)\,G_i,
\end{equation}
where $G_i$ is an effective thermal conductance relating a small probe-temperature change to a radiative power variation. The parameters $G_i$ are determined by matching the measured far-field flux to the SCUFF-EM prediction, where coupling-induced corrections are negligible. This effective calibration captures the conductive leak through the cantilever, the finite thermal participation of the cantilever itself and residual differences between the dielectric function used in the calculations and that of the actual borosilicate spheres. By convention, $\Delta Q_{\mathrm{rad},i}>0$ corresponds to an increase of the net radiative heat received by sphere~$i$ when the separation is reduced from $d_\mathrm{ref}$ to $d$. Error bars combine type-A and type-B uncertainty contributions. The shaded bands around the SCUFF-EM curves represent the uncertainty on the gap distance. At the smallest separations, this distance uncertainty dominates because residual lateral misalignment becomes comparable to the nominal gap.

\subsection{Dielectric function and isolated-sphere emissivity}
\label{Mie}

\begin{figure}[!h]
    \centering
    \includegraphics[width=\linewidth]{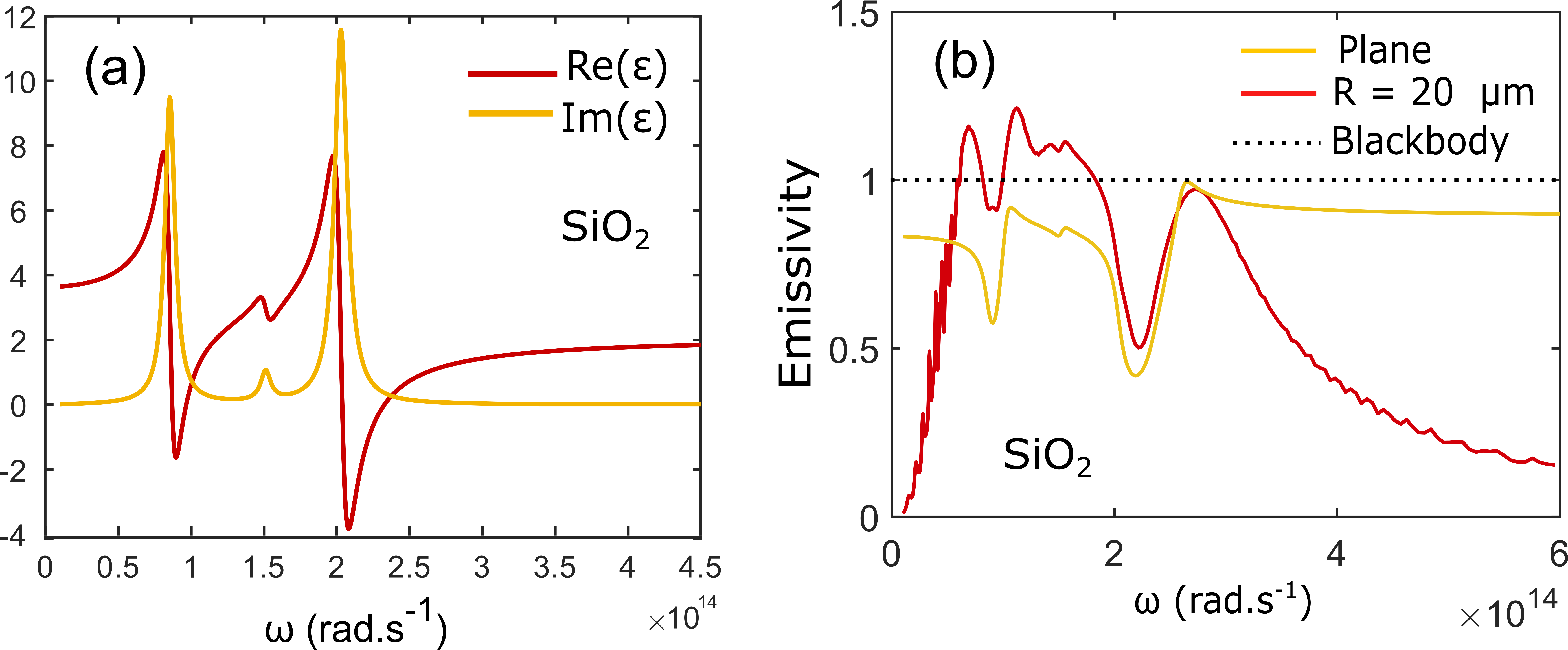}
    \caption{\textbf{Dielectric constant and emissivity of the glass microsphere} (a) Real and imaginary part of the dielectric function of silica as a function of the frequency. (b) Emissivity of a $R = $ \SI{20}{\micro\meter} silica spheres calculated in the framework of fluctuational  electrodynamics according to~\cite{kattawar_radiation_1970}. }
    \label{fig:Supp_mie}
\end{figure}

The dielectric function used throughout this work corresponds to that of silica and is shown in Fig.~\ref{fig:Supp_mie}(a), where both the real and imaginary parts are plotted as a function of frequency. Although the microspheres used in the experiment are made of borosilicate glass, we approximate their dielectric response by that of silica, which provides a good description of their infrared optical properties. The same dielectric function is used consistently in both the semi-analytical model and the full-wave numerical calculations performed with \textsc{SCUFF-EM}.

The emissivity of an isolated sphere is calculated using the framework developed by Kattawar and Eisner in their work ``Radiation from a Homogeneous Isothermal Sphere''~\cite{kattawar_radiation_1970}. In this approach, Rytov's fluctuational electrodynamics is applied to a spherical geometry to obtain the spectral power emitted by a sphere of radius $R$ and dielectric function $\varepsilon(\omega)$. The spectral radiative power emitted at angular frequency $\omega$ reads

\begin{equation}
    \Phi(\omega) = 4 \pi R^2 \frac{\omega^3 \hbar}{4 \pi^2 c^2} 
    \left[ \frac{1}{2} + \frac{1}{e^{\frac{\hbar \omega}{k_B T}} - 1} \right] 
    \hat{e}(\omega)\, \mathrm{d}\omega,
    \label{eqn:kattawar}
\end{equation}

where $\hat{e}(\omega)$ denotes the spectral emissivity of the sphere. The optical response of the sphere is described using Mie theory. The scattering and absorption cross-sections are defined as the ratio of the scattered or absorbed power to the incident irradiance on the sphere. The extinction cross-section satisfies

\begin{equation}
    \sigma_\mathrm{ext} = \sigma_\mathrm{sca} + \sigma_\mathrm{abs}.
    \label{eqn:sigma_ext}
\end{equation}

Dimensionless efficiencies are introduced as

\begin{equation}
    Q_\mathrm{i} = \frac{\sigma_\mathrm{i}}{\pi R^2},
\end{equation}

where $i$ denotes scattering, absorption, or extinction. These efficiencies can be expressed as infinite series over the multipole order $n$, with the size parameter $x = kR$. The scattering efficiency reads

\begin{equation}
    Q_\mathrm{sca}= \frac{2}{x^2} 
    \sum_{n=1}^{\infty} (2n+1)\left(|a_n|^2 + |b_n|^2\right),
\end{equation}
while the extinction efficiency is
\begin{equation}
    Q_\mathrm{ext}= \frac{2}{x^2} 
    \sum_{n=1}^{\infty} (2n+1) \Re(a_n + b_n).
\end{equation}

The absorption efficiency is therefore

\begin{equation}
    Q_\mathrm{abs}= \frac{2}{x^2} 
    \sum_{n=1}^{\infty} (2n+1) 
    \left[\Re(a_n + b_n) - |a_n|^2 - |b_n|^2 \right].
    \label{eqn:Q_abs}
\end{equation}

The Mie coefficients $a_n$ and $b_n$ are given by

\begin{equation}
    a_n = \frac{\psi_n(x)}{\zeta_n(x)} 
    \frac{D_n(y) - m D_n(x)}{D_n(y) - m G_n(x)},
\end{equation}

\begin{equation}
    b_n = \frac{\psi_n(x)}{\zeta_n(x)} 
    \frac{m D_n(y) - D_n(x)}{m D_n(y) - G_n(x)},
\end{equation}

where $y = mx$ with $m=\sqrt{\varepsilon(\omega)}$ the complex refractive index. The functions $\psi_n(x)$ and $\zeta_n(x)$ are Ricatti–Bessel functions, and $D_n$ and $G_n$ denote their logarithmic derivatives (see the full derivation in~\cite{kattawar_radiation_1970}).

Figure~\ref{fig:Supp_mie} shows the emissivity calculated for a silica sphere of radius $R=\SI{20}{\micro\meter}$. In a certain frequency range ($\omega \approx \SI{0.9e14}{\radian\per\second}$ to $\SI{1.8e14}{\radian\per\second}$), the emissivity exceeds unity. This behavior originates from the definition of emissivity for finite objects, which is normalized by the geometrical cross-section rather than the total emitting surface.

Using this Mie-based emissivity, the radiative flux emitted by an isolated sphere can be calculated. We verify that this approach reproduces exactly the isolated-sphere flux obtained from the \textsc{SCUFF-EM} simulations used in the main text, ensuring consistency between the semi-analytical description and the full-wave numerical calculations.

\subsection{Semi-analytical model}
\label{semi-analytical}

To provide a physically transparent interpretation of the measured fluxes, we introduce a semi-analytical description in which each sphere exchanges radiation both with the surrounding thermal bath and with the other sphere. Within this framework, the net radiative power received by sphere $i$ can be decomposed into three contributions,

\begin{equation}
Q_{\mathrm{rad},i}(d)=
Q_{\mathrm{bath} \rightarrow i}^{\mathrm{FF}}(d)
+Q_{j\rightarrow i}^{\mathrm{FF}}(d)
+Q_{j\rightarrow i}^{\mathrm{NF}}(d),
\end{equation}

where the three terms correspond respectively to far-field exchange with the surrounding bath, far-field exchange with the other sphere, and near-field exchange between the two spheres. The relative importance of these contributions depends strongly on the separation distance $d$. At large separations, the near-field term vanishes and radiative exchange is dominated by propagating modes, whereas at short distances the evanescent contribution becomes significant.

Within fluctuational electrodynamics, the radiative flux can be written in Landauer form as

\begin{multline}
Q_{\mathrm{rad},i}(d)=
\int_{0}^{\infty}\frac{\mathrm{d}\omega}{2\pi}
\Big[
\Delta\Theta(\omega,T_i,T_0)\,
\mathcal{T}_{\mathrm{bath} \rightarrow  i}^{\mathrm{FF}}(\omega,d)
\\
+\Delta\Theta(\omega,T_i,T_j)\,
\mathcal{T}_{j\rightarrow i}^{\mathrm{FF}}(\omega,d)
+\Delta\Theta(\omega,T_i,T_j)\,
\mathcal{T}_{j\rightarrow i}^{\mathrm{NF}}(\omega,d)
\Big],
\end{multline}

where $\Delta\Theta(\omega,T_i,T_j)=\Theta(\omega,T_j)-\Theta(\omega,T_i)$ and 
$\Theta(\omega,T)=\hbar\omega/(e^{\hbar\omega/k_\mathrm{B}T}-1)$ is the mean energy of the Planck oscillator.

\paragraph*{Far-field radiative exchange.}

In the far-field regime we employ a radiometric description based on the view factor $F(d)$ between two identical spheres~\cite{campbell_radiant-interchange_nodate}. This model does not account for interferences, multiple reflection and wave related effects. Within this approximation, the far-field transmission coefficients read
\begin{align}
\mathcal{T}_{j\rightarrow i }^{\mathrm{FF}}(\omega,d)
&=
2R^2\,\varepsilon^2_{\mathrm{sphere}}(\omega)\,
\frac{\omega^2}{c^2}\,F(d),
\\
\mathcal{T}_{\mathrm{bath} \rightarrow i}^{\mathrm{FF}}(\omega,d)
&=
2R^2\,\varepsilon_{\mathrm{sphere}}(\omega)\,
\frac{\omega^2}{c^2}\,[1-F(d)],
\end{align}
where $\varepsilon_{\mathrm{sphere}}(\omega)$ is the emissivity of an isolated microsphere computed above. In this radiometric picture, the radiation emitted by a sphere is partitioned geometrically between the bath and the neighboring sphere according to the view factor. The sum of the two contributions therefore equals the total far-field emission of an isolated sphere.

\paragraph*{Near-field contribution.}

The near-field transmission coefficient $\mathcal{T}_{j\rightarrow i}^{\mathrm{NF}}$ is evaluated using the Derjaguin approximation for sphere--sphere interactions~\cite{Derjaguin,otey_numerically_2011}, which relates the sphere--sphere problem to the plane--plane transmission coefficient calculated exactly within fluctuational electrodynamics~\cite{joulain_surface_2005}. The transmission coefficient reads
\begin{equation}
\mathcal{T^\mathrm{NF}_\mathrm{sub}} (\omega,d)
=
\int_{0}^{R}
\mathrm{d}r\;
\mathcal{\tilde{T}^\mathrm{NF}_\mathrm{sub}}
(\omega,\tilde{d}(r))\,2\pi r ,
\end{equation}
with the local separation between infinitesimal rings of the spheres
\begin{equation}
\tilde{d}(r)=
d+2R-2\sqrt{R^2-r^2}.
\end{equation}

The plane–plane evanescent transmission coefficient is

\begin{equation}
\mathcal{\tilde{T}^\mathrm{NF}_\mathrm{sub}}(\omega,d)
=
\sum_{s,p}
\int_{\omega/c}^{\infty}
\frac{\mathrm{d}\kappa}{2\pi}
\frac{4\kappa
\mathrm{Im}(r^{s,p}_{13})
\mathrm{Im}(r^{s,p}_{23})}
{|1-r^{s,p}_{12}r^{s,p}_{13}e^{-2ik_\perp d}|^2}
e^{-2\mathrm{Im}(k_\perp)d}.
\end{equation}

\begin{figure}[!h]
    \centering
    \includegraphics[width=\linewidth]{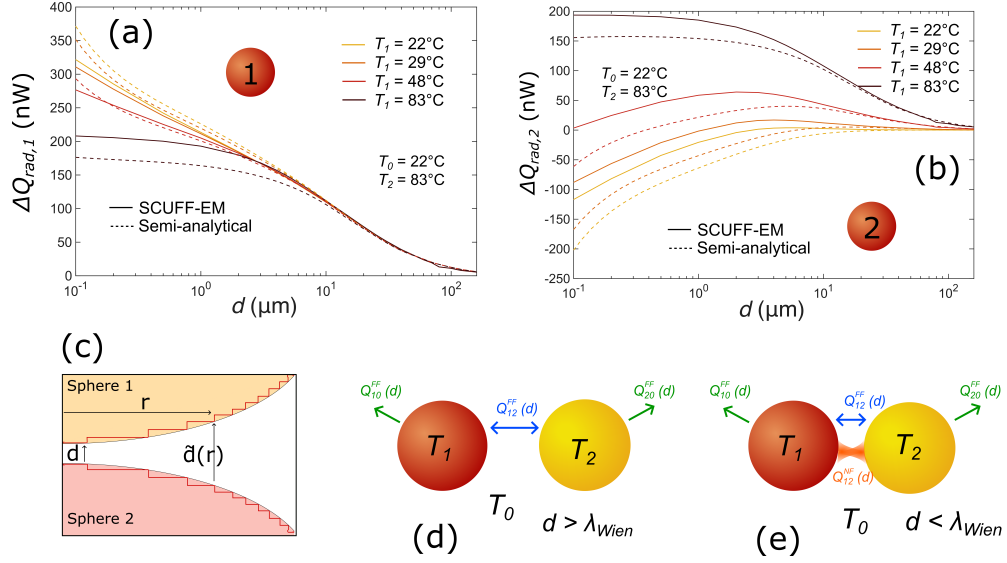}
    \caption{\textbf{Comparison between the semi-analytical model and full-wave calculations.}
    (a,b) Comparison of the modulated radiative flux predicted by the semi-analytical model and by \textsc{SCUFF-EM}. The model captures the qualitative behavior of the system, including the asymmetry between the two spheres and the non-monotonic evolution of the flux of the hotter sphere. 
    (c,d) Illustration of the different radiative channels at large and small separations. At large distances radiative exchange is dominated by propagating modes, whereas at short separations evanescent near-field coupling becomes significant.}
    \label{fig:Supp_semi}
\end{figure}

\paragraph*{Validity and limitations of the model.}

Figure~\ref{fig:Supp_semi} compares the predictions of the semi-analytical model with full-wave calculations performed with the boundary-element solver \textsc{SCUFF-EM}. The semi-analytical model reproduces the main qualitative features of the measurements, including the asymmetry between the spheres and the non-monotonic behavior of the radiative flux.

However, quantitative deviations appear when the separation becomes comparable to or smaller than the thermal wavelength ($\lambda_{\mathrm{Wien}}\sim\SI{10}{\micro\meter}$). In this regime the radiometric description becomes insufficient because it neglects multiple scattering, interference effects and the modification of the electromagnetic environment produced by the nearby sphere. These wave effects are fully captured by the \textsc{SCUFF-EM} calculations and are responsible for the renormalization of the radiative emission discussed in the main text.

The semi-analytical model should therefore be regarded primarily as a physically transparent decomposition of the different radiative channels rather than a quantitatively exact description at the smallest separations.

\subsection{Determination of the dressed emissivity}
\label{dressed}

In order to quantify how the presence of a nearby object modifies the far-field emission of the two-sphere system, we introduce the concept of dressed emissivity. This quantity describes how the apparent emissivity of a sphere is modified by its electromagnetic environment.
We consider the variation of the radiative power received by each sphere when the separation distance $d$ between the spheres is changed, relative to a reference configuration at large separation $ d_\mathrm{ref} $. The modulated radiative power received by sphere $i$ is defined as
\begin{equation}
\Delta Q_{\mathrm{rad},i}(T_0,T_1,T_2;d)=
Q_{\mathrm{rad},i}(T_0,T_1,T_2;d)-Q_{\mathrm{rad},i}(T_0,T_1,T_2;d_{\mathrm{ref}}),
\end{equation}

The net radiative power received by sphere $i$ results from two contributions: the radiation exchanged with the other sphere and the radiation exchanged with the external bath at temperature $T_0$. It can therefore be written as
\begin{equation}
Q_{\mathrm{rad},i}(T_0,T_i,T_j;d)
= Q_{j \rightarrow i}(T_j,T_i;d) + Q_{\mathrm{bath} \rightarrow i }(T_0,T_i;d),
\end{equation}
where $Q_{j\rightarrow i}$ denotes the power radiated by sphere $j$ and absorbed by sphere $i$, while $Q_{\mathrm{bath}\rightarrow i}$ represents the power received by sphere $i$ from the surrounding environment.

To characterize the external emission of the coupled system, we introduce a dressed emissivity $\varepsilon^{\mathrm{dress}}(d)$ such that the radiative power emitted by each sphere toward the bath can be expressed as
\begin{equation}
Q_{\mathrm{bath} \rightarrow i}(T_0, T_i;d)=
A\,\sigma\,\varepsilon^{\mathrm{dress}}(d)\left(T_0^4 - T_i^4\right),
\end{equation}
where $A=4\pi R^2$ is the surface area of a sphere and $\sigma$ is the Stefan--Boltzmann constant. Because the radiative exchange between the two spheres conserves energy, the mutual power exchange satisfies
\begin{equation}
Q_{1\rightarrow2}(T_1, T_2;d) = - Q_{2\rightarrow1}(T_1,T_2;d).
\end{equation}
As a consequence, when summing the modulated powers received by the two spheres the mutual sphere--sphere contribution cancels, and only the bath-sphere contributions remain, yielding
\begin{equation}
\Delta Q_{\mathrm{rad,tot}}(T_0,T_1,T_2;d)
= \Delta Q_{\mathrm{bath}\rightarrow 1}(T_0,T_1;d) + \Delta Q_{\mathrm{bath}\rightarrow 2}(T_0,T_2;d),
\end{equation}
which corresponds to the net variation of radiative exchange between the two-sphere system and the surrounding bath. From this relation, the dressed emissivity variation can be directly obtained as
\begin{equation}
\Delta \varepsilon^{\mathrm{dress}}(d)
=
\frac{\Delta Q_{\mathrm{rad,tot}}(T_0,T_1,T_2;d)}
{A\sigma\left(2T_0^4 - T_1^4 - T_2^4 \right)},
\end{equation}
where $\Delta \varepsilon^{\mathrm{dress}}(d)=
\varepsilon^{\mathrm{dress}}(d) - \varepsilon^{\mathrm{sphere}}$ and $\varepsilon^{\mathrm{sphere}}$ is the exact emissivity of an isolated sphere calculated above.

This formulation highlights that the dressed emissivity directly captures how the photonic environment created by the nearby sphere modifies the far-field thermal emission of the system. In particular, deviations from the isolated-sphere emissivity arise from near-field electromagnetic coupling, which alters the electromagnetic modes available for thermal radiation(LDOS) and thus affects the thermal flux radiating through the surrounding bath.

\end{document}